\documentclass[12pt,a4paper]{article}
\usepackage[numbers]{natbib}
\usepackage{rotating}
\usepackage{graphicx}

\begin{document}

\begin{center}
{\Large \bf Simple Standard Model Extension by Heavy Charged Scalar
\hfill\\}
\end{center}
\vspace{0.5cm}

\begin{center}
{E.~Boos$^{1,2}$, I.~Volobuev$^{1,2}$\\
\hfill\\
{\small \it $^1$Skobeltsyn Institute of Nuclear Physics, Lomonosov Moscow State University}\\
{\small \it  Leninskie Gory, 119991, Moscow, Russia} \\
 {\small \it $^2$Faculty of Physics, Lomonosov Moscow State University,
 Leninskie Gory}\\
{\small \it 119991, Moscow, Russia}  }
\end{center}

\begin{abstract}
We consider a Standard Model extension by a heavy charged scalar
gauged only under the $U_{Y}(1)$ weak hypercharge gauge group.
Such an extension, being gauge invariant with respect to the SM
gauge group, is a simple special case of the well known Zee model.
Since the interactions of the charged scalar   with the Standard
Model fermions turn out to be significantly suppressed compared to
the Standard Model interactions, the charged scalar provides an
example of a long-lived charged particle being interesting to
search for at the LHC. We present the pair and single production
cross sections of the  charged scalar at different colliders and
the possible decay widths for various boson masses. It is shown
that the current ATLAS and CMS searches at 8 and 13 TeV collision
energy lead to the bounds on the scalar boson mass of about
300--320 GeV. The limits are expected to be much larger for higher
collision energies  and, assuming $15~ab^{-1}$ integrated
luminosity, reach about 2.7 TeV at future 27 TeV LHC thus covering
the most interesting mass region.
\end{abstract}

\section{Introduction}
With the discovery of the Higgs boson at the LHC, the Standard
Model (SM) was completed in the sense that all the predicted
particles have been found and all the interaction structures have
been fixed. However, not all the interactions in the gauge and
Higgs sectors are confirmed experimentally. The Standard Model is
based on the fundamental principles such as gauge invariance, the
absence of chiral anomalies, unitarity and renormalizability. It
is a common knowledge that the SM works extremely well explaining
an enormous amount of experimental facts and results. However,
because of a number of theoretical problems such as the hierarchy
problem and the inability to explain  the presence of Dark Matter
or the nature of CP violation, the SM is considered as a sort of
effective theory describing phenomena up to the electroweak or TeV
energy scale. A large number of various experimentally allowed
 beyond the SM models and scenarios are proposed motivating
intensive searches for new physics in the terrestrial and space
experiments, in particular, at the LHC. However, up to now no
convincing results confirming any concrete BSM direction have been
obtained.

Among various objects predicted by new physics models a special
attention has been recently paid to the so-called HSCP (heavy
stable charged particles) or LLP (long-lived particles). Various
SM extensions predict the existence of such particles
\cite{Farrar:1978xj}-\cite{Huitu:2010uc}. A number of searches for
LLP and HSCP have been performed at the Tevatron and the LHC
\cite{Abazov:2008qu}-\cite{Khachatryan:2016sfv}.

In this paper we discuss shortly a very simple SM extension by a
charged scalar boson interacting with the $U_{Y}(1)$ weak
hypercharge gauge boson and  potentially giving an example of a
long-lived charged particle. Such a model from rather different
perspectives has been considered  in paper \cite{Bilenky:1993bt}
and quite recently in paper \cite{Cao:2017ffm}. This SM extension
by the extra charged scalar can be naturally called csSM.

Generic SM extensions by an arbitrary number of  Higgs singlets
and doublets were considered by P. Langacker in his famous review
paper \cite{Langacker:1980js}. We consider in more detail one
particular case with an extra complex scalar field $S$ interacting
in a gauge invariant manner  only with the $U_{Y}(1)$ weak
hypercharge  gauge field and with the Higgs field. The scalar
field potential of the  model coincides with that of the SM
extension by singlet complex  scalar with $U(1)$ symmetry
discussed in paper \cite{Barger:2008jx}, where this scalar field
couples only to the Higgs field and is shown to give a reliable
explanation of the cold dark matter. In our model we identify this
$U(1)$ symmetry with the weak hypercharge $U_{Y}(1)$ symmetry,
which makes the complex scalar electrically charged and forbids
its interpretation as a dark matter particle. The model (csSM) can
be viewed as a simplified variant of the Zee model
\cite{Zee:1980ai}. The  original Zee model includes  an extra
scalar $SU(2)$ doublet and gives rise to a number of
intriguing interactions in the lepton sector, which lead to
processes with lepton number violation \cite{Zee:1980ai} (see a
recent discussion in \cite{Thao:2017qtn}), and  to radiatively
induced Majorana neutrino masses \cite{Zee:1985id, Babu:1988ki}.
The parameter space of the Zee model allowed by the experimental
data has been recently worked out \cite{Herrero-Garcia:2017xdu}
showing that the masses of the additional scalars in the range of
a few hundreds GeV are possible, but they have to lie in the range
below a few TeV.

\section{The Minimal Model}
The minimal part of the   SM  Lagrangian extended by the scalar
field carrying a non-trivial representation of the $U_{Y}(1)$ weak
hypercharge group includes the terms of dimension not greater than
four. If one requires, in addition,  lepton number conservation,
as it takes place in the SM, the simplest model Lagrangian
contains the kinetic term and the mass and self-coupling terms of
the charged scalar boson field:
\begin{equation}
L_{S} = D_{\nu}^{*}S^* D^{\nu}S - V(S)\, ,
\label{Lgrn}
\end{equation}
where the covariant derivative is given by $D_{\nu} =
\partial_{\nu} - i g_1 \frac{Y_S}{2} B_{\nu}$,
 $B_{\nu}$ being the SM weak hypercharge gauge field, $g_1$ is the SM $U_{Y}(1)$ coupling and
 $Y_S$ is the  weak hypercharge of the new scalar field $S$.

The potential  $V(S)$ may have, in general, the following  gauge
invariant form
\begin{equation}
V(S) = \mu_S^2 |S|^2 + \lambda_{S} (|S|^2)^2 + \lambda_{\Phi S}|\Phi|^2|S|^2\, ,
\label{Potential}
\end{equation}
where $\mu_S^2$ is a mass parameter, $\lambda_{S}$  is the
S-scalar  quartic self-coupling, $\lambda_{\Phi S}$ is the
coupling of the S-scalar to the Higgs field, which is supposed to
be less than one in order to keep the model in the perturbative
regime. The last term has been included into the potential,
because it contributes to the mass term after spontaneous symmetry
breaking.

Let us stress a few points here:
\begin{itemize}
\item {The S-field is a charged field, so it cannot have a nontrivial vacuum expectation value.
Therefore, it cannot influence the value  the  SM  $\rho$ -
parameter.}
\item {Since the gauge boson $B$ is expressed in the SM as a linear combination of the photon
and the Z-boson fields, $B_{\nu} = A_{\nu} cos\theta_W - Z_{\nu}
sin\theta_W$, the S-scalar couples to the photon with the constant
$e\frac{Y_S}{2}$, where the electromagnetic constant $e$  is equal
to  $g_1 cos\theta_W$, as it is usual in the SM. The S-scalar is
an electrically charged field. As will be shown later, the
hypercharge of the S-field is equal to two with the electric
charge being  equal to one ($Q_S=Y_S/2$). Thus, we denote the
$S$-field as $S^-$ and the complex conjugate field $S^*$ as $S^+$.
}
\item
{The model described by  Lagrangian (\ref{Lgrn}) has three
physically relevant parameters: the charged scalar mass squared
$M_S^2 = \mu_S^2 + 1/2\lambda_{\Phi S}v^2$ which has to be
positive; the coupling $\lambda_{\Phi S}$, which is assumed to
satisfy the condition $|\lambda_{\Phi S}|<1$  corresponding to the
perturbative regime; and the positive self-interaction coupling
$\lambda_{S}$. The mass term parameter $\mu_S$ can be equal to
zero. In this case the mass of the S-boson comes from the
interaction with the Higgs field in a similar way as for the other
SM particles  and is equal to $M_S^2 = \lambda_{\Phi S}v^2/2 $.
Then  its natural value is of the order of hundred GeV, which, as
it will be discussed below, is ruled out by the present LHC data.
}
\item
{If only the  dimension 4 or less operators are included, there
are no gauge invariant operators containing the charged scalar and
the quark fields. We did not include into the Lagrangian the gauge
invariant operators of  dimension four, which describe the
interaction of the S-scalar with the SM lepton fields giving
lepton number violating vertices, they will be discussed shortly
later. As a result, in this approximation the S-scalar is a stable
particle. }
\end{itemize}
In a simplest variant of the model the last property leads  to the
prediction of a stable charged scalar boson. Obviously, if the
mass of the boson is of the order of a few hundreds GeV, the
existence of the boson will not contradict the limits from
precision electroweak measurements, in particular, the limits on S
and T-parameters \cite{Herrero-Garcia:2017xdu}.  However, an
important question  is, whether  the existence of such particles
is compatible with bounds from cosmology.  The production and
freeze-out of S-scalars would be similar to those of cold dark
matter particles and can be described by the same formulas
\cite{Gorbunov-Rubakov}. If we consider the particle to be stable,
the bounds come from the restrictions on the abundance of such
scalars. Since positively charged S-scalars could form
super heavy hydrogen,   the abundance of S-scalars should be
much less than the relative abundance of tritium, otherwise the
S-scalar would have been already discovered in natural water.
Estimates with the help of the micrOMEGAs program
\cite{Belanger:2018ccd} show that, for the S-scalar mass 200 GeV
and larger, it is possible in scenarios with low reheating
temperature of the order of $ 4$ GeV. In this case the ratio of
the abundance of S-scalars to that of hydrogen is approximately
$10^{-23}$, which is five orders of magnitude less than the
relative abundance of tritium about  $10^{-18}$
(see, \cite{tritium}). The latest  direct 
searches for super heavy (or anomalously heavy) hydrogen in deep sea water at 
4000~m taking into account gravitational concentration
gradients  give the upper limit
for the relative abundance of such particles about
$4\times10^{-17}$ in the mass range of  5~GeV -- 1.6~TeV  at  95\%
confidence level \cite{Yamagata:1993jq}.

Negatively charged S-scalars could form bound states with protons and
deuterons, which could catalyze nuclear fusion \cite{Ioffe:1979tv}. However, the
binding energy of these states would be of the order of 50\, KeV,
and they could not exist during Big Bang nucleosynthesis (BBN). If
the S-scalar can decay, the situation is quite different and will
be briefly discussed in section 4.

\section{Pair Production Cross Sections}
\begin{figure}[h]
\centerline{
\includegraphics[width=.8\textwidth]{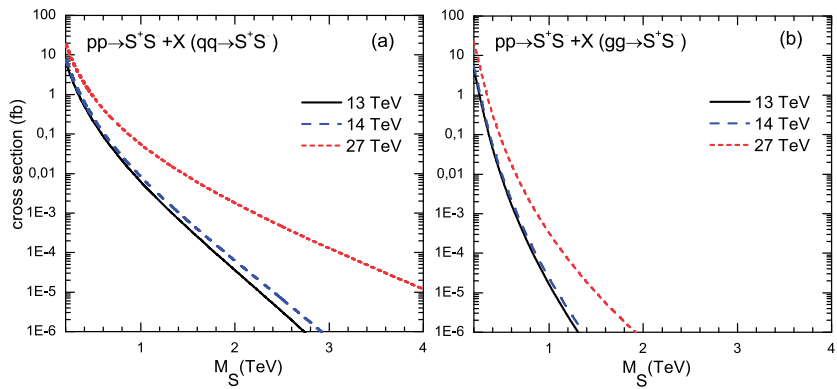}}
\caption{Charged scalar pair production cross section via photon
and Z-boson exchange in quark anti-quark annihilation (a) and via
Higgs boson exchange in gluon-gluon fusion (b) at $\sqrt{s}=13,
14, 27$~TeV as a function of its mass.} \label{fig:1}
\end{figure}

Charged scalars can be produced at the LHC  in pairs via the
Drell-Yan process in collisions of quark-antiquark pairs as well
as in the gluon-gluon fusion. The production cross section as a
function of the charged scalar mass is shown in Fig.\ref{fig:1}
for three different proton-proton collision energies $\sqrt{s} =
13, 14, 27$~TeV. \footnote{The computations here and below have
been performed by means of the CompHEP program \cite{CompHEP},
into which the Feynman rules obtained from the Lagrangian under
consideration by means of the LanHEP code \cite{LanHEP}  were
implemented.} One can see from Fig.\ref{fig:1} that the cross
section grows with the collider energy. For the Drell-Yan process
initiated by the quark-antiquark collisions the cross section is
about $10~fb$ for 200 GeV mass for the energy 13, 14 TeV. More
accurately the LO cross section is about $6.8~fb$ and $7.6~fb$ at
13 TeV and 14 TeV respectively with the NNLO K-factor for the
Drell-Yan quark-antiquark type of processes (Fig.\ref{fig:1}a) of
about 1.18 \cite{Hamberg:1990np, Anastasiou:2003yy, Catani:2009sm}
. The cross section rapidly goes down with  the increase of the
scalar mass. For example, for 1 TeV scalar mass the cross section
is about $5 \times 10^{-2}$~fb even for the energy 27 TeV expected
for high energy (HE) regime of the LHC operation. This would lead
to the production of a few hundreds charged scalar pairs in the
case of a very high luminosity of about $15~ab^{-1}$.  The
production in quark-antiquark pair collisions was also discussed
in paper \cite{Cao:2017ffm}.

 There is an additional contribution to the pair
production cross section, which comes from the gluon-gluon fusion
mechanism and was not discussed in \cite{Cao:2017ffm}. Two gluons
produce a virtual SM Higgs boson via the top loop triangle diagram
and the virtual Higgs boson decays to a pair of the charged
scalars. The production cross section as a function of the mass of
the S-boson for the maximal boundary value of the coupling
constant $\lambda_{\Phi S} = 1$, is shown in Fig.\ref{fig:1}b for
the collision energies $\sqrt{s}=13, 14, 27$~TeV. For smaller
values of the coupling the cross section has to be just multiply
by the factor $\lambda_{\Phi S}^2$. For the computation we have
used the well-known expression for the triangle top loop diagram
which was specially implemented into the CompHEP code.  The cross
section in Fig.\ref{fig:1}b includes the NNLO K-factors as given
in \cite{Harlander:2002wh}. Since the scalar boson production via
the gluon-gluon fusion is described by exactly the same diagram as
the resonant SM Higgs production the K-factor is the same as for
the SM Higgs with  the Higgs mass equal to $2\times{M_S}$. The
gluon-gluon fusion cross section very rapidly decreases with the
grows of the scalar boson mass. The level of the cross section is
comparable with that for the Drell-Yan quark-antiquark
annihilation process in Fig.\ref{fig:1}a only for small masses
close to 200 GeV and it becomes significantly smaller than the
quark-antiquark annihilation part even for the maximum value of
the coupling $\lambda_{\Phi S} = 1$. Therefore, the gluon-gluon
production mode gives practically negligible contribution to the
pair production rate.

 Searches for stable charged particles at the LHC energy 13
TeV presented in \cite{Khachatryan:2016sfv} give  the lowest bound
on the production cross section of about $4~fb$ -- $2~fb$ for the
luminosity $2.5~fb^{-1}$ corresponding to 10 -- 5 events as the
lowest number of events expected for the stable charged particle
production.  This bound is found for the case of the stau leptons
having the same pair production mechanism as the S-scalars. From
the bound one gets the lower limit on the charged scalar mass of
about 270 GeV and 300 GeV using the cross section given in
Fig.\ref{fig:1}.  The cross section limit about $0.5~fb$ obtained
in RUN1 at the LHC energy 8 TeV and much higher luminosity
$18.8~fb^{-1}$ for the case of the stau lepton
\cite{Chatrchyan:2013oca} leads to a slightly stronger lower
S-scalar mass limit of about 320 GeV due to smaller cross section
at 8 TeV than at 13 TeV.

Assuming the same lowest number of expected events from 10 to 5
one can estimate from the computed cross sections the expected
lower limits on the boson mass for various cases of collision
energies and luminosities. So, for the proton-proton collision
energy 14 TeV and the luminosity $300~fb^{-1}$ the expected mass
limits are calculated to be about 800 GeV and 950 GeV
respectively. For the benchmark energy 27 TeV and the luminosity
$15~ab^{-1}$ the limits on the charged scalar mass are expected to
be 2.4 TeV and 2.7 TeV.

\begin{figure}[h]
\centerline{
\includegraphics[width=.4\textwidth]{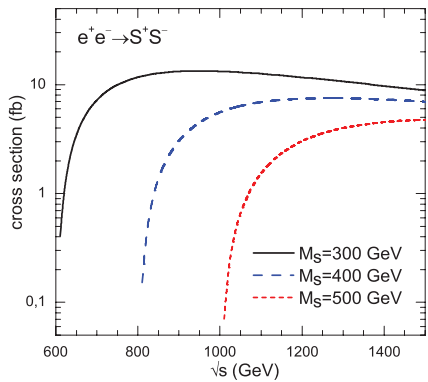}}
\caption{Charged scalar pair production cross section in $e^+ e^-$
collisions  as a function of collision centre of mass energy for
the scalar mass $M_S = 300, 400, 500$ GeV.} \label{fig:2}
\end{figure}

For completeness the production cross section in $e^+ e^-$
collisions  is shown in Fig.2 as a function of collision energy
for the scalar mass 300 GeV, 400 GeV, and 500 GeV. The level of
the cross section in Fig.\ref{fig:2} is large enough giving good
prospects to study  the charged scalars in detail, if its mass is
in the kinematically accessible range.  However,  the scalar in
that mass range having the  specified production cross sections in
hadronic collisions (Fig.\ref{fig:1}) will be, most probably,
either ruled out or, in the case of luck, discovered at the LHC
before a $e^+ e^-$ linear collider a large enough energy will
start to operate.

\section{Interactions with Leptons and Quarks}
If  only the above discussed  terms (operators) of dimension 4 had
been present in the extended SM Lagrangian, the charged scalar
boson would  not have had interactions leading to its decay and/or
single production, and therefore the boson would have been stable.
However, gauge invariant operators of dimension four and five
involving the charged scalar field can be constructed, which lead
to decays of the boson. We will first discuss the gauge invariant
terms of dimension four involving the lepton fields.

 The left-handed lepton  doublet ${l}_a$ of each generation
$a=1,2,3$ carries the representation $\underline 2 (-1)$ of the
group $SU(2)\times U_Y(1)$. The conjugate doublets ${\bar l}_a$
and the charge conjugate doublets ${l}_b^c$ transform as
$\underline 2^* (1)$. Since the scalar field $S^-$ carries the
representation $\underline 1 (-2)$ of this group,   the dimension
4 gauge invariant terms ${\bar l}_a\epsilon\,{l}_b^c S^-$  can be
constructed, $\epsilon$ denoting the standard antisymmetric
$2\times2$ matrix with $\epsilon_{12} =1$. These terms give rise
to the coupling of the S-scalar to leptons, which are
antisymmetric in the generation indices due to the matrix
$\epsilon$. Therefore, the transformation properties of the
S-scalar field under the gauge group of the SM allow the existence
of the following interactions, which can be explicitly written as
\cite{Zee:1980ai}:
\begin{equation} \label{SC_lepton_inter}
 L_{\mathrm{S,leptons}} =\left( f_{12}(\bar\mu_L \nu_e^c - \bar e_L \nu_\mu^c) +
 f_{13}(\bar\tau_L \nu_e^c - \bar e_L \nu_\tau^c) + f_{23}(\bar\tau_L \nu_\mu^c - \bar \mu_L
 \nu_\tau^c)\right)S^- + h.c.,
 \end{equation}
where $\nu^c$ denotes the charge conjugate neutrino field. These
interactions lead to lepton flavor violation as well as to
violation of the lepton number by two units due to the involvement
of the charge conjugate fields.
 However, it turns out that at low
energies this lepton number violation is very small due to the
large S-scalar mass. Moreover, one can show with the help of Fierz
identities that the S-scalar mediated interactions of leptons
conserve lepton number and  can be brought to the standard form of
Fermi's four fermion interaction, which imposes  constraints on
the coupling constants $f_{ik}$
\cite{McLaughlin:1999rr,Mituda:2001jw}. The results of these
papers with the present day values of the Fermi constant
\cite{vanRitbergen:1999fi, pdg} and the probabilities of the
decays $\tau \rightarrow \mu \bar \nu_\mu \nu_\tau,\, \tau
\rightarrow e \bar \nu_e \nu_\tau,\, \mu \rightarrow e \gamma$
\cite{pdg} give  $|f_{12}|^2 < 3\times 10^{-6} G_F M_S^2$,
$|f_{13}|^2, |f_{23}|^2  < 2.8\times 10^{-2} G_F M_S^2$. A full
parameter scan of the Zee model carried out in paper
\cite{Herrero-Garcia:2017xdu} and including a fit of the neutrino
mixing angles and  mass differences gives the constraints on the
coupling constants $f_{ik}$, which  turn out to be much more
stringent: $|f_{12}|, |f_{13}|,|f_{23}| < 10^{-6}$.  For these
values of the coupling constants the partial widths of the
S-scalar decays to leptons are less than 0.5~eV for the S-scalar
mass up to 5~TeV.

The interaction of the S-scalar with the quark fields can take
place only due to  gauge invariant terms of dimension five or
larger. Here we will  discuss the gauge invariant terms of
dimension five involving the quark fields. To introduce the
notations let us first recall the well-known fact that, in the SM,
the most general interaction Lagrangian of the Higgs field and the
quark fields includes a mixing of the fermion fields from various
generations:
 \begin{equation}
 L_{\mathrm{Yukawa}} = -\Gamma_d^{ij}\bar {Q'_{\rm L}}^i\Phi {d'_{\rm R}}^j  -\Gamma_u^{ij}\bar {Q'_{\rm L}}^i\Phi^C {u'_{\rm R}}^j
+ \mathrm{h.c.},
 \label{SM_quarks}
 \end{equation}
where $\Gamma_{u,d}$ are generically possible mixing coefficients
with up- and down-type quark fields. The Higgs and the conjugate
Higgs $SU_L(2)$ doublet fields in the unitary gauge are
 $$\Phi=\frac{1}{\sqrt{2}}
 \left(
\begin{array}{c}
0\\v+h
\end{array}
\right)\; \mbox{ and }\;\Phi^C=i\sigma^2\Phi^{\dag}=\frac{1}{\sqrt{2}}
 \left(
\begin{array}{c}
v+h\\0
\end{array}
\right)$$
After spontaneous symmetry breaking  Lagrangian
(\ref{SM_quarks}) in the unitary gauge takes the following form
\begin{equation}
L_{\mathrm{Yukawa}} =
- \left[M_d^{ij} \bar{d'_{\rm L}}^i {d'_{\rm R}}^j + M_u^{ij} \bar{u'_{\rm L}}^i{u'_{\rm R}}^j
 + \mathrm{h.c.}
\right]\cdot\left(1+\frac{h}{v}\right),
 \label{L_yukawa_quarks}
 \end{equation}
where $  M^{ij} = \Gamma^{ij}v/{\sqrt 2}$ is a generic mass mixing matrix.

In order to obtain the physical mass  eigenstates of quarks, the
matrices $M^{ij}$ should be diagonalized by unitary
transformations of the left- and right-handed quark fields:
 \begin{equation}
 d'_{\mathrm{L}i} = (U_{\rm L}^d)_{ij}d_{\mathrm{L}j};\,\,\,\,\,d'_{\mathrm{R}i} = (U_{\rm R}^d)_{ij}d_{\mathrm{R}j};\,\,\,\,\,u'_{\mathrm{L}i}
= (U_{\rm L}^u)_{ij}u_{\mathrm{L}j};\,\,\,\,\,u'_{\mathrm{R}i} = (U_{\rm R}^u)_{ij}u_{\mathrm{R}j}
\label{mixing_q}
\end{equation}
\begin{equation}
U_{\rm L}^{u,d} (U_{\rm L}^{u,d})^{\dag} =1,\;\;\;U_{\rm R}^{u,d} (U_{\rm R}^{u,d})^{\dag} =1.
\label{unitarity}
\end{equation}

The matrices $U$ are chosen such that
$$(U^u_{\rm L})^{\dag}M_u U_{\rm R}^u=\left(
\begin{array}{ccc}
m_u&0&0\\0&m_c&0\\0&0&m_t
\end{array}
\right);\,\,\,\,\,(U^d_{\rm L})^{\dag}M_d U_{\rm R}^d=\left(
\begin{array}{ccc}
m_d&0&0\\0&m_s&0\\0&0&m_b
\end{array}
\right)$$
 As it is well known, the SM neutral currents remain the
same after the above unitary transformation providing the absence
of the FCNC at three level. However,  after the transformation to
the physical degrees of freedom
$$u' \rightarrow (U^u_L)u, \, \, \,
d' \rightarrow (U^d_L)d,
$$
the charged currents get a unitary
matrix in front of the down quark fields,
$$
 V_{CKM} =(U^u_L)^{\dag} U^d_L,
$$
called the Cabbibo-Kobayashi-Mascawa (CKM) mixing matrix.
Similarly, after the unitary transformation of the lepton fields,
one gets the Pontecorvo-Maki-Nakagawa-Sakata neutrino mixing
matrix (PMNS) in front of the massive neutrino fields in the
charged leptonic currents.

In a similar manner one can write a gauge invariant Lagrangian for
the interaction of the SM fermions with the charged scalar boson:
\begin{equation}
 L_{\mathrm{S,quarks}} = -\frac{1}{\Lambda}\bar {Q'_L} \lambda_u \Phi {u'_R}^j S^-  -
 \frac{1}{\Lambda}\bar {Q'_L}\lambda_d\Phi^C {d'_R} S^+ + \mathrm{h.c.},
 \label{SC_quarks_inter}
 \end{equation}
where $\lambda_{u,d}$ are dimensionless matrices  and
$\Lambda$ is the scale of "new physics". After the substitution of
the Higgs field and the transformation (\ref{mixing_q})
of the quark fields to  the mass eigenstates,  one gets the following
interaction Lagrangian in the unitary gauge
\begin{equation}
 L_{\mathrm{S,quarks}} =
- \frac{1}{\Lambda}\left[\bar{d} \cdot V_u \frac{1+\gamma_5}{2}
 u \cdot S^- \, + \,\bar{u} \cdot V_d \frac{1+\gamma_5}{2}
 d \cdot S^+  + \mathrm{h.c.}
\right]\cdot\left(1+\frac{h}{v}\right)
 \label{L_unit}
\end{equation}
where $V_{d} = V_{CKM}(U^d_{\rm L})^{\dag}\mu_d U_{\rm R}^d$ and
$V_{u} = V_{CKM}^{\dag} (U^u_{\rm L})^{\dag}\mu_u U_{\rm R}^u$,~~
$\mu_{d,u} = \lambda_{d,u} v/{\sqrt 2}$. The elements of matrices
$\mu_{d,u}$ have the dimension of mass, the matrices are not
diagonal in general, they may contain complex phases leading to CP
violation. Here we do not discuss such a general case.

If we assume that the Minimal Flavour Violation hypothesis
\cite{Chivukula,DAmbrosio:2002vsn} is valid, the matrices
$\mu_{d,u}$ are proportional (or equal) to the mass matrices
$M^{ij}$. In this case the matrices $V_{d,u}$ are equal to the
products of the CKM matrix or its hermitian conjugated matrix and
the diagonal mass matrices for the up- and down-type quarks. The
interactions of the two first quark generations are therefore
naturally suppressed by the corresponding quark masses allowing to
overcome the FCNC constrains \cite{Herrero-Garcia:2017xdu}. The
dominating part is the interaction of the charged scalar with the
top-bottom quark charged current. In fact, the interaction
structure is very similar to that of the charged Higgs in the 2HDM
or MSSM taken at $\tan\beta$=1 (see \cite{Gunion:1984yn,
Gunion:1989we, Branco:2011iw, Akeroyd:2016ymd, Dev:2014yca}).
However, in comparison with the 2HDM or MSSM the interaction
vertices are suppressed by the factor of the order of $v/\Lambda$.

 It is worth noting that interactions similar to those
described by formulas (\ref{SC_quarks_inter}) and (\ref{L_unit})
can exist also in the lepton sector. If the neutrinos are
considered to be massless, the corresponding formulas will include
only the terms similar to the second ones in formulas
(\ref{SC_quarks_inter}), (\ref{L_unit}). If the neutrinos are
considered to be massive, they will be absolutely similar to
formulas (\ref{SC_quarks_inter}), (\ref{L_unit}). However, it is
natural to expect the entries of  the corresponding mass matrices
$\mu_{\nu,e} = \lambda_{\nu,e} v/{\sqrt 2}$ to be of the order of
neutrino and charged lepton masses, and in this case the
contribution of these terms to the S-scalar decay processes is
negligible compared with the decay to t-quark.

The dominating production channel $pp \rightarrow t + S^- +X$ in
the case of the scalar boson being heavier than the top quark is
similar to the charged Higgs case with the suppression factor
$(v/\Lambda)^2$. If the scale is not very large, the production
cross section could be large enough to be interesting for searches
at the LHC as shown in Fig.3.
 \begin{figure}[h]
\centerline{
\includegraphics[width=.4\textwidth]{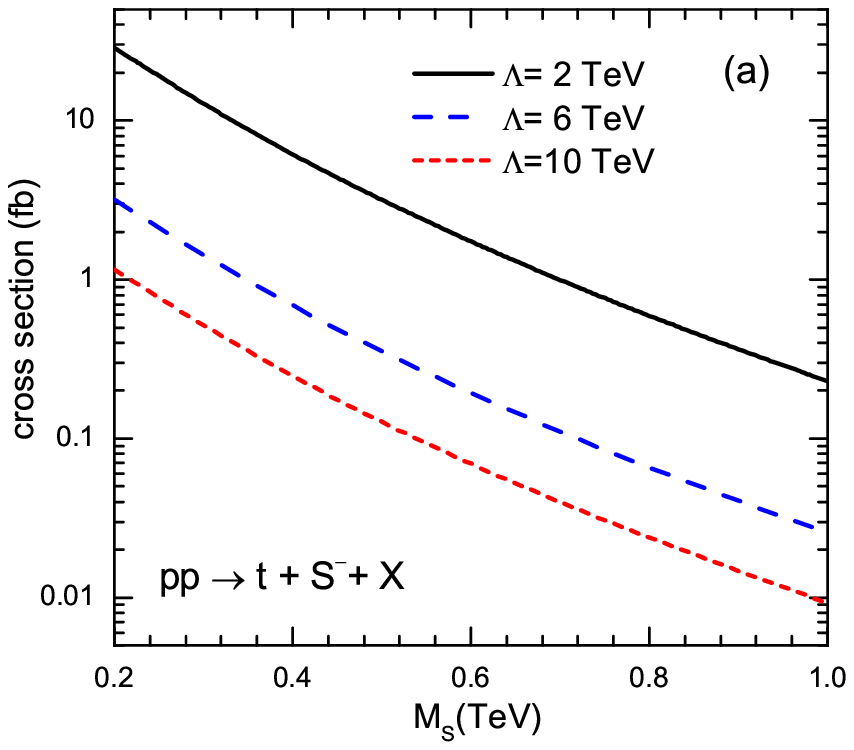}
\includegraphics[width=.4\textwidth]{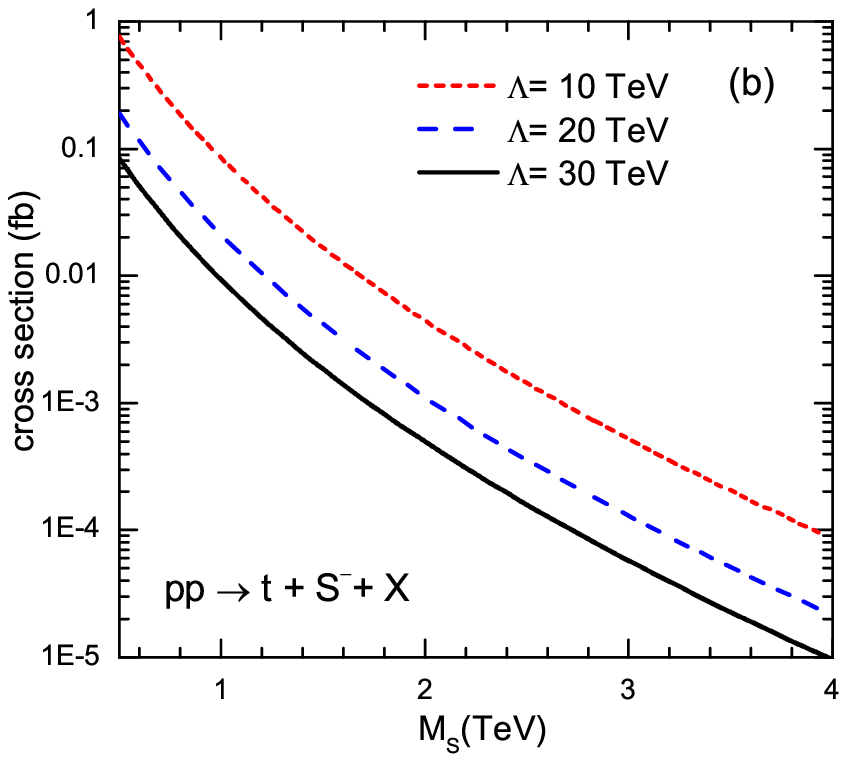}}
\caption{Charged scalar single production cross section at 14 TeV
(a) and at 27 TeV (b) pp collision energy as a function of the
scalar mass for three values of the scale $\Lambda$ 2 TeV, 6 TeV,
and 10 TeV.} \label{fig:3}
\end{figure}
The NLO corrections make the result much more stable with respect
to the factorization/renormalization scale variation while the NLO
K-factor is found to vary in the range of 1.4 or less
\cite{Dittmaier:2009np}. The single production cross section
decreases quadratically with the scale and becomes smaller than
the considered above pair production at some scale.
 \begin{figure}[h]
\centerline{
\includegraphics[width=.4\textwidth]{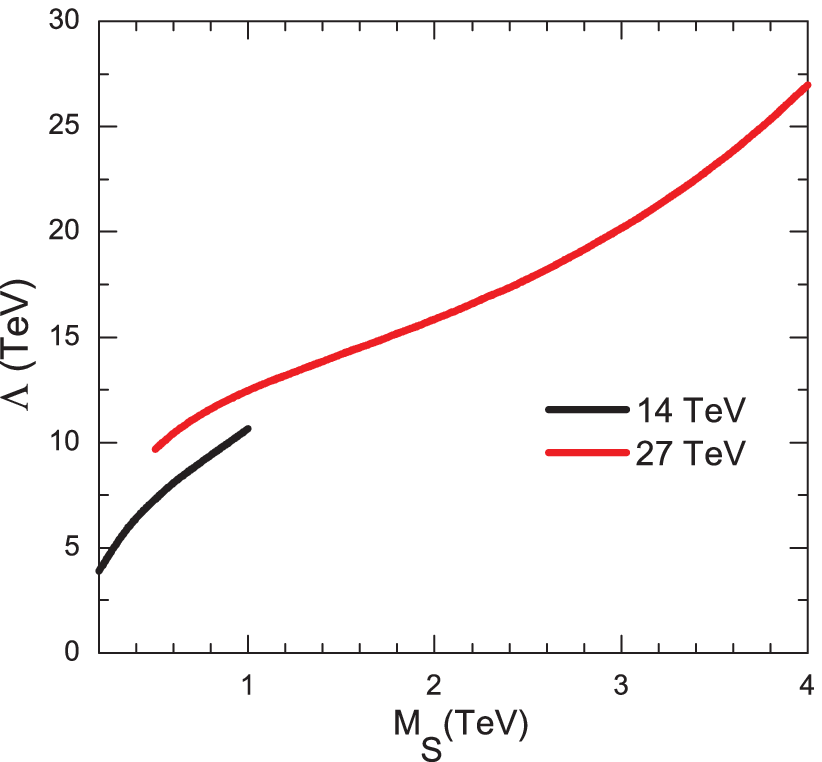}}
\caption{The scale $\Lambda,$ at which the single production cross
section becomes equal to the pair production cross section at 14
TeV and at 27 TeV pp collision energy, as a function of the scalar
mass.} \label{fig:5}
\end{figure}
In Fig.\ref{fig:5} the dependence of the scale $\Lambda$,
 corresponding to equal single and pair charged scalar
 production cross sections, is shown for 14 TeV
and 27 TeV collision energy as a function of the scalar boson mass
for the the same mass intervals as in Fig.\ref{fig:3}. For the
values of the scale above the lines in Fig.\ref{fig:5} the single
production cross section is smaller than the pair one and vice
versa, for the values below the lines the single production cross
section is larger.

\begin{figure}[h]
\centerline{
\includegraphics[width=1.0\textwidth]{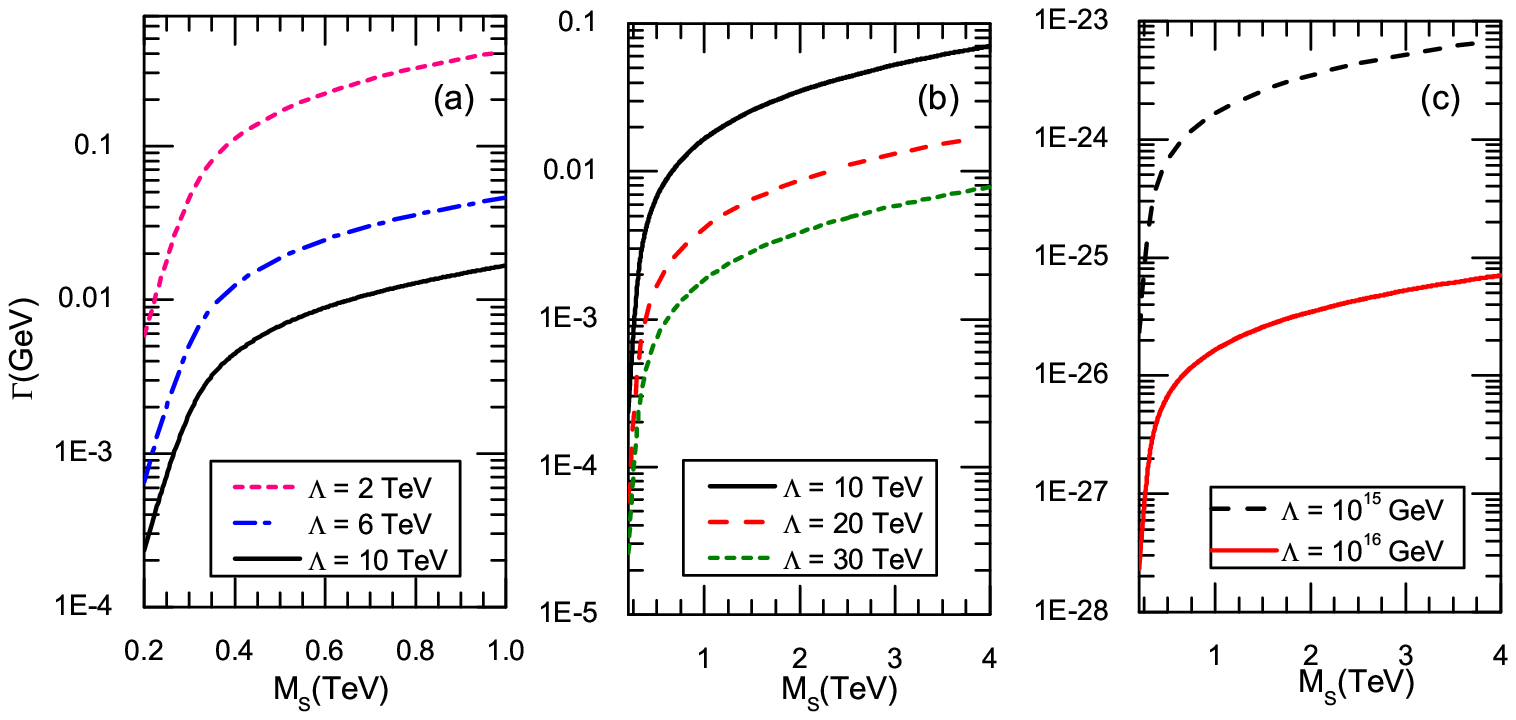}}
\caption{Charged scalar width as a function of its mass at new
physics scale in TeV range (a),(b) and in GUT range (c).}
\label{fig:4}
\end{figure}

As was mentioned, the mass of the S-scalar in the csSM may arise
from the SM Brought-Engler-Higgs (BEH) mechanism. In this case the
natural values for the scalar mass would be in the range of the
Higgs vacuum expectation value. The scale $\Lambda$ may originate
from completely different physics and could be much larger. The
S-scalar decays to the top-bottom pair with nearly 100\%
probability.  The decay width is proportional to
$M_{top}^2/\Lambda^2\cdot M_S\cdot \beta^3$ ($\beta =
\sqrt{1-M_{top}^2/M_S^2}$) and therefore increases with the scalar
boson mass and rapidly decreases with the growth of the scale
$\Lambda$. This is demonstrated in Fig.\ref{fig:4}
One can see that for the energy scale $\Lambda$ in TeV range the
scalar boson decay width varies from $10^{-1}$ GeV to $10^{-4}$
GeV (left plot). However if the scale is in the GUT range (right
plot) the width becomes  very small  $10^{-24}$ GeV --$10^{-27}$
GeV. In this case the life time of the scalar might be $0.1$ sec
or more leading to a microscopic travel distance before the decay.
For the case of large scales  $\Lambda$ the single boson
production cross section becomes negligible at colliders, and the
charged scalars may be produced only in pairs.  This corresponds
to the case of long-lived charged particles with  the discussed
above current  and expected limits on the charged scalar boson
mass.   The existence   of such particles  is compatible with
cosmology, if they decay before BBN. The micrOMEGAs program
\cite{Belanger:2018ccd} gives that this is the case, if $\Gamma >
10^{-24}$ GeV. Fig. \ref{fig:4} shows that, for the S-scalar mass
200 GeV and larger, the width satisfies this restriction for all
the considered values of $\Lambda$,  except $\Lambda = 10^{15}$
GeV for the S-scalar mass smaller than about 700 GeV and $\Lambda
= 10^{16}$ GeV for all the considered S-scalar masses.

 For rather small energy scales  $\Lambda$ in TeV range the
charged scalar may be produced either singly or in pairs with
subsequent decays into top and bottom quarks. However, both
production cross sections are significantly smaller than the top
pair and the single top cross sections. Topologies involving the
top and bottom quarks in the final states have been discussed to
be rather promising to search for the charged Higgs boson with
mass heavier than the top mass (see the study in
\cite{Akeroyd:2016ymd}).  Corresponding searches have been
performed recently \cite{ATLAS:2016qiq} and the limits on the
production cross section times the decay branching to the top and
bottom was found to be about 1 pb for the charged Higgs mass about
300 GeV  and  down to about 0.2 pb for the mass range around 1
TeV.  For the case of $\tan \beta =1$ in the 2HDM or MSSM, where
the interaction of the charged Higgs boson is very similar to that
of the S-scalar,  the lower limit on the charged Higgs boson mass
was found to be about 400 GeV. From Fig.\ref{fig:1} and
Fig.\ref{fig:3} one can see that the cross sections for the
S-scalar production in both pair and single production channels
are expected to be significantly smaller.  Therefore, in this case
a special analysis is needed in order to estimate, whether or not
a small signal of the charged scalar could be extracted from much
larger backgrounds at the LHC, in particular, in high energy and
high luminosity operation regimes.

\section{Concluding Remarks}
A simple gauge invariant extension of the SM considered in this
study may provide an example of a heavy stable charged (HSCP) or
long-lived (LLP) particle. The model contains, in addition to the
SM fields, only the charged scalar field gauged only under the
$U_{Y}(1)$ weak hypercharge gauge group. The model can be
considered as a simple special case of the well known Zee model.
In the simplest case, assuming the presence of only dimension 4
operators and  lepton number conservation, the gauge invariant
Lagrangian of the model contains only the gauge interaction of the
charged scalar and its interaction with the SM Higgs field. Since
in this case one cannot construct  gauge invariant interactions of
the scalar with the SM fermions, the charged scalar boson is a
stable particle. The main production mode is the charged scalar
Drell-Yan pair production via the photon and Z-boson exchange in
quark-antiquark and via the SM Higgs exchange in gluon-gluon
collisions. From the computed cross sections and the results of
searches for HSCP at the LHC one can estimate the current bounds
on the charged scalar boson mass to be about 300-320 GeV and the
expected bounds at higher collision energies and larger
luminosity. In particular, at future 27 GeV LHC with the
luminosity of $15~ab^{-1}$ the bound is expected to reach about 2.7 TeV
covering a significant part of interesting mass regions following from the
overall parameter space analysis for the Zee model as found in
\cite{Herrero-Garcia:2017xdu}.

Allowing higher dimensional
operators and  violation of the lepton number one can add to the
Lagrangian the interactions of charged scalar field with the SM
fermions leading to decays of the scalar boson. The dimension 4
operators containing lepton fields violate lepton number
conservation, and the corresponding coupling strengths are
significantly constrained by the muon decay, the neutrino mass
measurements and oscillation data. The dimension 5 operators  in
the quark sector are naturally proportional to the fermion masses
and the CKM matrix elements. The dominating decay mode of the
charged scalar boson is, therefore, the decay to the top and the
bottom quarks and the dominating single boson production channel
is the associated production with the top quark. This is rather
similar to the charged Higgs production and decay in 2HDM or MSSM
at $\tan{\beta}=1$, although  with an additional suppression by
the factor $\frac{v^2}{\Lambda^2}$. The single production cross
section varies from $10~fb$ to $10^{-5}~fb$ in the mass range
between 200 GeV and  4 TeV and in the range of the scale $\Lambda$
from 2 TeV to 30 TeV. The decay width depends strongly on the
scalar boson mass and the scale $\Lambda$ and for the TeV scale
regions takes values from 100 MeV to 0.1 MeV or so. If the scale
is much larger, say, in the GUT range, the decay width to the top
and bottom quarks becomes very small. In this case the width could
be dominated by lepton number violating decays, but this obviously
depends on the small lepton violating coupling strengths.

The estimates made with the help of the micrOMEGAs program
\cite{Belanger:2018ccd} show  that the existence of such a heavy
stable   or long-lived  particle is not forbidden by cosmology in
some regions of the parameter space. A more detailed study of
cosmological consequences of this model will be done in a separate
paper.

\section{Acknowledgements}
We  thank Viatcheslav Bunichev, Eduard Rahmetov, and Tatiana
Tretyakova for useful discussions. Special thanks go to
Alexander Pukhov for interesting discussions and  help with
calculations using the micrOMEGAs code. We are grateful to the
Russian Science Foundation (grant 16-12-10280) for  support.

\end{document}